\documentclass[preprint,12pt]{elsarticle-mod}

\usepackage{amssymb}

\usepackage{microtype}

\usepackage[hyphens]{url}

\usepackage{amsmath}

\usepackage{color}
\usepackage{longtable}
\usepackage{overpic}
\usepackage{graphicx}

\begin{document}

\begin{frontmatter}



\title{Terraforming the dwarf planet: \\ Interconnected and growable Ceres megasatellite world}


\author[FMI,Aurora,Tartu]{Pekka Janhunen\corref{cor1}}
\ead{pekka.janhunen@fmi.fi}
\ead[url]{http://www.electric-sailing.fi}

\address[FMI]{Finnish Meteorological Institute, Helsinki, Finland}
\address[Aurora]{Also at Aurora Propulsion Technologies Oy, Espoo, Finland}
\address[Tartu]{Also at University of Tartu, Estonia}
\cortext[cor1]{Corresponding author}

\begin{abstract}
We analyse a megasatellite settlement built from Ceres materials in
high Ceres orbit. Ceres is selected because it has nitrogen, which is
necessary for an earthlike atmosphere. To have $1 g$ artificial
gravity, spinning habitats are attached to a disk-shaped megasatellite
frame by passively safe magnetic bearings. The habitats are
illuminated by concentrated sunlight produced by planar and parabolic
mirrors. The motivation is to have a settlement with artificial
gravity that allows growth beyond Earth's living area, while also
providing easy intra-settlement travel for the inhabitants and
reasonably low population density of 500 /km$^2$. To enable gardens
and trees, a 1.5 m thick soil is used. The soil is upgradable to 4 m
if more energy is expended in the manufacturing phase. The mass per
person is $10^7$ kg, most of which is lightly processed radiation
shield and soil. The goal is a long-term sustainable world where all
atoms circulate.  Because intra-settlement travel can be
propellantless, achieving this goal is possible at least in
principle. Lifting the materials from Ceres is energetically cheap
compared to processing them into habitats, if a space elevator is
used. Because Ceres has low gravity and rotates relatively fast, the
space elevator is feasible.
\end{abstract}


\end{frontmatter}



\pagebreak[4]

\section*{Nomenclature}
\nobreak\noindent
\begin{longtable}{ll}
$A$          & Cross-sectional area \\
au           & Astronomical unit, 149\,597\,871\,km \\
$E$          & Energy of burn (Appendix) \\
$F$          & Force \\
$G$          & Gravitational constant, $6.674\cdot 10^{-11}$ Nm$^2$/kg$^2$ \\
$g$          & Earth's acceleration of gravity, 9.81 m/s$^2$ \\
$g_x, g_y$   & Gravitational field components \\
$k$          & Numeric coefficient (Appendix) \\
$L$          & Generic angular momentum (Appendix) \\
$L_m$        & Angular momentum due to lifting of mass $m$ (Appendix) \\
$L_C$        & Angular momentum of Ceres \\
$m$          & Mass of lifted propellant (Appendix) \\
$M$          & Mass of lifted payload (Appendix) \\
$M_C$        & Mass of Ceres, $9.38\cdot 10^{20}$ kg \\
$r$          & Orbit radius, $10^5$ km \\
$r_0$        & Radial distance of tether tip (Appendix) \\
$r_1$        & Radial distance of tip of tether extension (Appendix) \\
$R$          & Radius of megasatellite disk \\
$R_C$        & Radius of Ceres, 470 km \\
$R_H$        & Radius of habitat, 1 km \\
$v_0$        & Thruster exhaust speed (Appendix) \\
$x,y$        & Coordinates \\
$Y$          & Compression \\
$\eta_T$     & Efficiency of the thruster (Appendix) \\
$\rho$       & Mass density \\
$\omega_C$   & Ceres angular rotation speed, $1.923\cdot 10^{-4}$ s$^{-1}$ \\
GCR          & Galactic cosmic radiation\\
ISRU         & In-situ resource utilisation \\
\end{longtable}\addtocounter{table}{-1} 

\section{Introduction}

Proposals for extraterrestrial settlements have generally fallen into two
categories. On one hand, there are plans to settle the surfaces of Moon
and Mars. A fundamental drawback is the
low gravity, which raises questions regarding long-term health effects,
especially for children whose muscles and bones are growing. Moon and
Mars also have smaller surface areas than Earth, so that
impact to the financial economy would be limited.

There have also been studies of spinning orbital settlements
\citep{ONeill1974,ONeill1977,AroraEtAl2006}. They have the benefit of
providing healthy $1 g$ artificial gravity. One of their challenges is
how to provide travel for goods and people between the free-flying
settlements. This issue arises as soon as the population grows larger
than what fits in one settlement unit. If the settlements are in
heliocentric orbit, they tend to drift apart. This would cause
travel times to grow to several months. This problem is avoided if the
settlements orbit a common body but then collision avoidance becomes
an issue. Lastly, for all orbits it holds that if travel is based on
rocket propulsion, the scheme is not long-term sustainable because
propellant atoms cannot be recycled. After exiting from the nozzle, a
propellant atom starts to orbit the Sun. Sooner or later it is ionised
by a solar ultraviolet photon, after which it is accelerated out from the
Solar System by the solar wind's electric field.

In this paper we study the possibility of attaching the spinning
habitats to a fixed frame, a megasatellite. Propellantless travel
between the habitats is then possible along the megasatellite's
structure. The geometry of the megasatellite is chosen in such a way
that the megasatellite is self-similarly growable. The intention is to
have only one megasatellite which grows as new settlements are
built. The rotary joints between the habitats and the megasatellite do
not limit the lifetime or cause reliability issues if one uses
passively safe Inductrack-type magnetic bearings with no physical
contact \citep{PostAndRyutov2000a,PostAndRyutov2000b}.  We choose
Ceres as the source body to be mined for the materials, because it has
nitrogen \citep{KingEtAl1992,DeSanctisEtAl2015}. Nitrogen is a
necessary component of the settlement's air.

\section{How people wish to live}

For the settlers to live well, the environment must be sufficiently
earthlike:
\begin{enumerate}
\item Earthlike radiation shielding.
\item Earthlike atmosphere.
\item $1 g$ artificial gravity.
\item 24 h diurnal cycle with 130 W/m$^2$ insolation, like in southern Germany.
\item Nature, fields, parks, forests.
\item Population density 500 /km$^2$, like in the Netherlands.
\item Large, interconnected world.
\end{enumerate}

The environment can also be better than Earth:
\begin{enumerate}
\item No adverse weather.
\item No natural disasters. (Concerning new threats, we
  show in subsection \ref{subsect:impactthreat} below that the
  meteoroid impact threat can be at least mitigated well or perhaps even eliminated.)
\item Ultimately growable to larger living area than Earth.
\end{enumerate}

The goal is that the artificial world is long-term
sustainable. The flux of escaping atoms must be small, and ideally all atoms
should circulate. One of the implications is that
daily use of in-space rocket propulsion is not acceptable, because propellant atoms escape from the
Solar System permanently. Even if an atom is not injected to Solar
System escape speed directly, eventually it gets ionised by a solar UV photon, after
which it is taken away by the solar wind.

One might ask if a reduced pressure pure oxygen atmosphere would also
work. It is known from the Gemini and Apollo programmes that
astronauts can live in such atmosphere without problems at least for a
few weeks. However, with our present knowledge, long term health
impacts to the lungs cannot be excluded. Even at reduced pressure, a
pure oxygen atmosphere also increases the risk of fire because there
is no inert gas that would contribute to cooling of a flame.  Because
the mass density of air would be low, it would also make it
challenging for insects and birds to fly. Insects are needed for
pollination, and people would miss birds if they were not
around. Hence we require an earthlike atmosphere with nitrogen.

We formulate the corresponding technical requirements:
\begin{enumerate}
\item Radiation shielding of 7600 kg/m$^2$, when made of 20\,\% of
  water and 80\,\% of silicate regolith, is sufficient. We made
  OLTARIS \citep{SingleterryEtAl2011} run with this composition, augmented by 1 kg/m$^2$ of
  boron-10 to absorb neutrons, for 1987 solar minimum conditions at 1
  au, with a spherical configuration which is the worst-case
  assumption, and obtained 20.45 mSv/year equivalent dose and 5.808
  mGy/year absolute dose. According to \citep{GlobusAndStrout2017}, 20
  mSv/year and 6 mGy/year are suitable limits.
\item The quality of artificial gravity is improved when the rotation
  radius is made larger. \citet{GlobusAndHall2017} analysed this issue
  and found that a rotation period of 30 s is a conservative choice
  and a period of 15 s or even 10 s gives acceptable quality.
  The question is fundamentally
  open, however, until crewed orbital experiments with artificial
  gravity are made. To be reasonably certain that quality of
  artificial gravity is high, we adopt a rotation radius
  of 1 km which corresponds to rotation period of 66 s.
\item Per person, there is 2000 m$^2$ of living space, which we divide
  into rural space (about 1100 m$^2$) and urban space (about 900
  m$^2$). The rural space is $1 g$ gravity. For the urban space we
  allow somewhat lower gravity, such as $0.8 g$.
\item To enable trees, we assume 1.5 m soil depth in the rural
  space. We assume soil density of 1500 kg/m$^3$, then the mass per
  unit area of the soil is 2250 kg/m$^2$. The rural space is
  illuminated by natural sunlight, with diurnally averaged insolation
  of 130 W/m$^2$.
\item The urban space has no soil and it uses artificial illumination.
\item The atmosphere height is 50 m and 15 m for rural and urban
  space, respectively.
\end{enumerate}

\section{Ceres as source body}

We choose Ceres as the source body from which the building materials
are mined.  The primary reason is that Ceres has nitrogen, which is
necessary for the settlement's atmosphere. Alternatively, one might
consider a carbonaceous asteroid, but those C-type asteroids that have
at least 100 km size and that have reasonably low eccentricities are
about equally far from Earth than Ceres. The asteroids are likely to
have less nitrogen than Ceres, although accurate data of Ceres and
C-type asteroid nitrogen abundance are not available at this point.

We choose to orbit Ceres so that the settlement remains physically
close to the source body. This ensures that the material transfer time
in the construction phase remains relatively short. We use a high
orbit to minimise tidal forces. We adopt $r=100,000$ km circular
equatorial orbit. This is smaller than the Hill's sphere radius of
Ceres (207,000 km), so that the orbit is expected to be long-term
stable without maintenance. We have not performed simulations to check
this, however, so this issue remains to be studied in the future.

We recommend using a space elevator to lift materials from Ceres. The
elevator cable length is 1024 km and the cable strength requirement is
straightforward to meet. Lifting from the surface takes only 54 kJ/kg
of energy. After elevator, we need 20 m/s of delta-v to circularise
the orbit.

Ceres is water-rich, so it would also be possible to make H$_2$/O$_2$
propellant and use rockets to lift the materials. We prefer the space
elevator, however, because it needs much less energy per lifted mass.

\section{Growing Ceres megasatellite}

\begin{figure}[htb]
\centering
\includegraphics[height=0.3\textwidth]{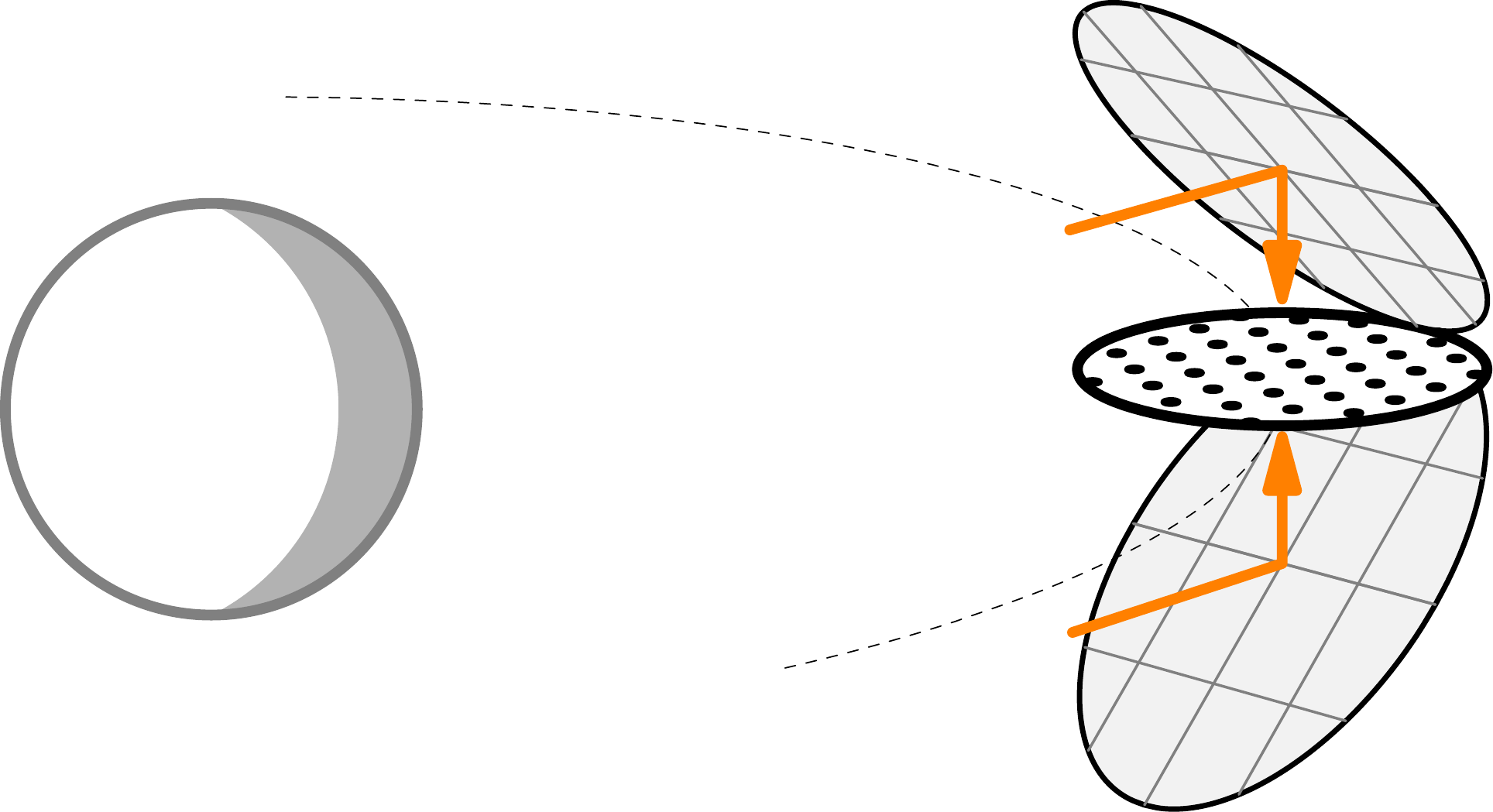}
\caption{Megasatellite in Ceres orbit.}
\label{fig:clamshell}
\end{figure}

Consider a disk-shaped megasatellite in equatorial Ceres orbit
(Fig.~\ref{fig:clamshell}). A number of spinning habitats are attached
to both sides of the disk, denoted by dots. To gather sunlight, two
45$^\mathrm{o}$ inclined mirrors are added. The megasatellite is in
microgravity so the mirrors and their support structures are
lightweight. Nearly all mass is in the habitats. Thus the mass
distribution can be made circularly symmetric so that the tidal torque
is zero. To keep the mass distribution symmetric, one can transfer
water or other ballast fluid between tanks. Any remaining tidal torque
must be treated by active attitude control, for example by using the
spinning habitats as reaction wheels.  Small changes in the level of
artificial gravity are not likely to be problematic for the
inhabitants. A propulsive attitude control system can exist as a
backup.

Because the orbit of Ceres has eccentricity, there is $\pm
8.7^\mathrm{o}$ libration about the correct Sun-pointing direction in
different parts of the orbit.  The libration occurs because the orbital
angular speed of Ceres varies according to Kepler's second law, while
the megasatellite rotates at a constant rate. One way to deal with the
libration would be to keep the megasatellite turned exactly towards the
Sun at all times by using large reaction wheels. A more attractive
solution is to make the mirrors from tiltable segments that are able
to direct sunlight vertically towards the disk even when sunlight
arrives a bit from the side. This reduces the amount of gathered light
by the cosine of the libration angle but the effect is insignificant
because worst-case factor is $\cos(8.7^\mathrm{o})=0.988$. The mirror
segments can be small and numerous, so the solution can be made
redundant and reliable. Thus, reaction wheels are not needed.

\begin{figure}[htb]
\centering
\includegraphics[height=0.3\textwidth]{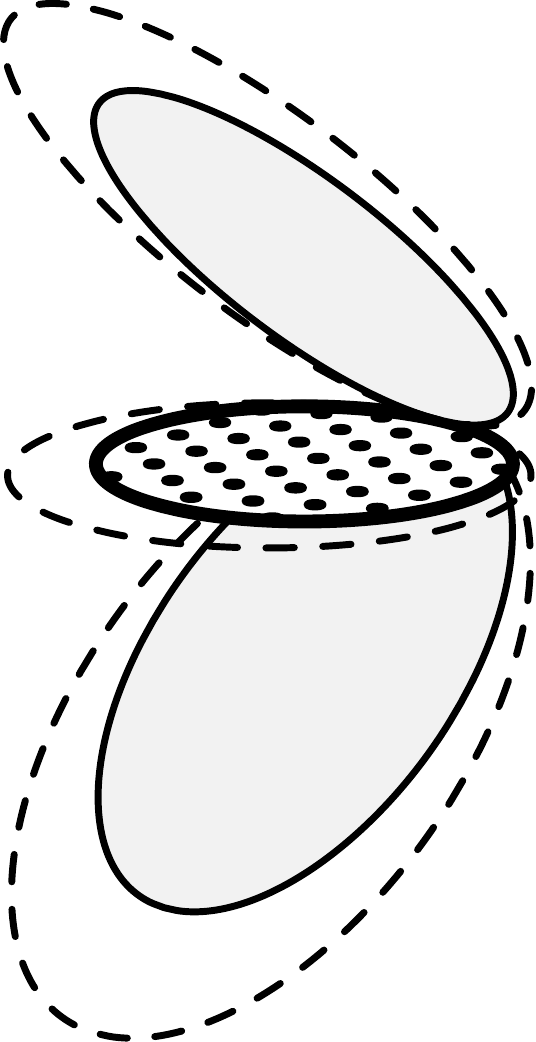}
\caption{Self-similar growth of the megasatellite.}
\label{fig:clamshellgrowth}
\end{figure}

The megasatellite of Fig.~\ref{fig:clamshell} is self-similarly
growable at the edges. One can extend the megasatellite by growing
both the disk and the mirrors as shown in
Fig.~\ref{fig:clamshellgrowth}.

\subsection{Gathering sunlight}

\begin{figure}[ht]
\centering
\begin{overpic}[width=0.5\textwidth]{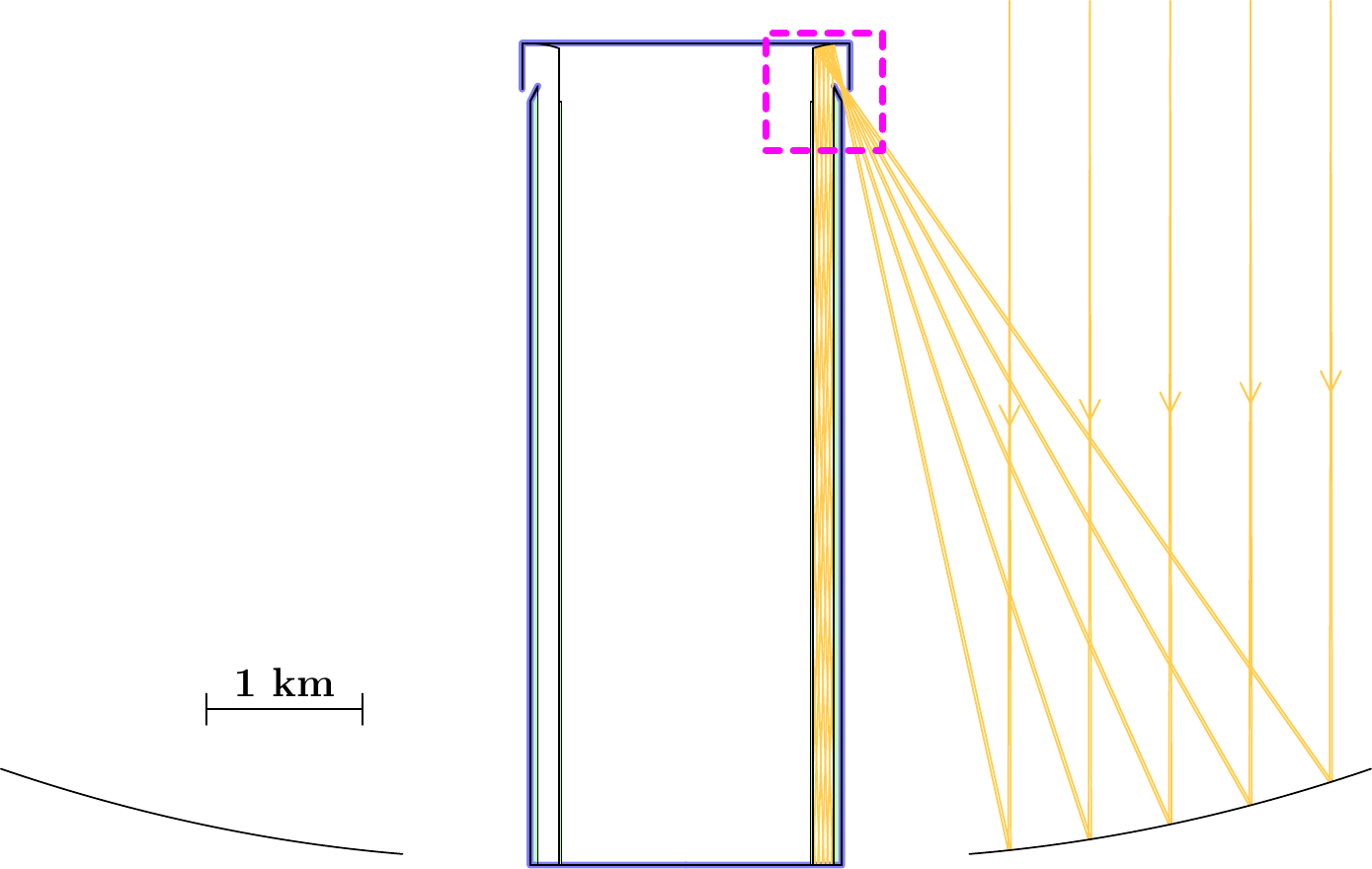}\put(3,50){a)}\end{overpic}
\quad\quad
\begin{overpic}[width=0.25\textwidth]{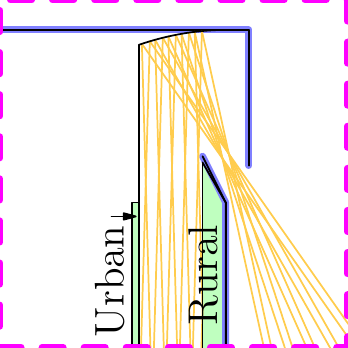}\put(3,70){b)}\end{overpic}
\caption{(a) Concentration of sunlight into rotating habitat, (b) detail.}
\label{fig:cylinder}
\end{figure}

Sunlight is concentrated into a rotating habitat by primary and
secondary mirrors, which are cylindrical paraboloids
(Fig.~\ref{fig:cylinder}). The primary mirror's focal ring coincides
with a light entry slit which is an opening in the habitat's radiation
shielding envelope. The opening is shaped such that the habited areas
do not have direct unshielded view to space. This is sufficient to
block radiation because high-energy radiation propagates straight
without bending around corners or appreciably reflecting from
surfaces: the straight-ahead approximation is typically valid in
shielding studies \citep{AlsmillerEtAl1968}, and the straight-ahead
approximation is also made in OLTARIS. The secondary mirror is inside
the envelope and it injects light into a light channel which is
parallel to the rotation axis.

\subsection{Synthetic diurnal cycle}

From the light channel, light is tapped to illuminate the rural space
through windows. The urban space is located inward of the light
channel and is artificially illuminated
(Fig.~\ref{fig:cylinder}b). The urban and rural space are concentric
cylinder surfaces.  The radius of the rural space is the same as the
habitat radius $R_H=1$ km.

\begin{figure}[ht]
\centering
\begin{overpic}[width=0.12\textwidth]{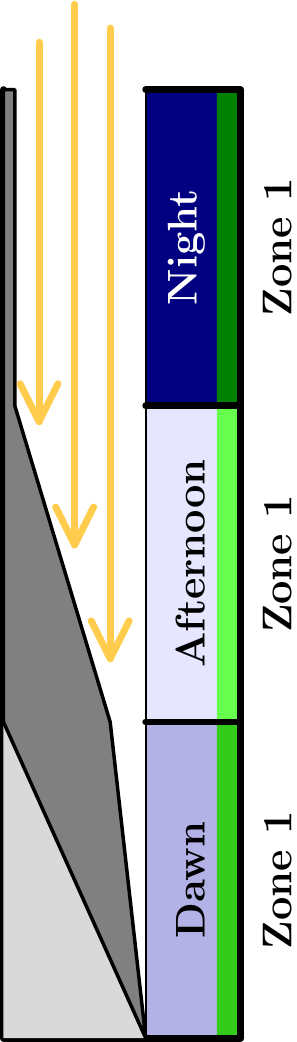}\put(3,85){a)}\end{overpic}
\begin{overpic}[width=0.63\textwidth]{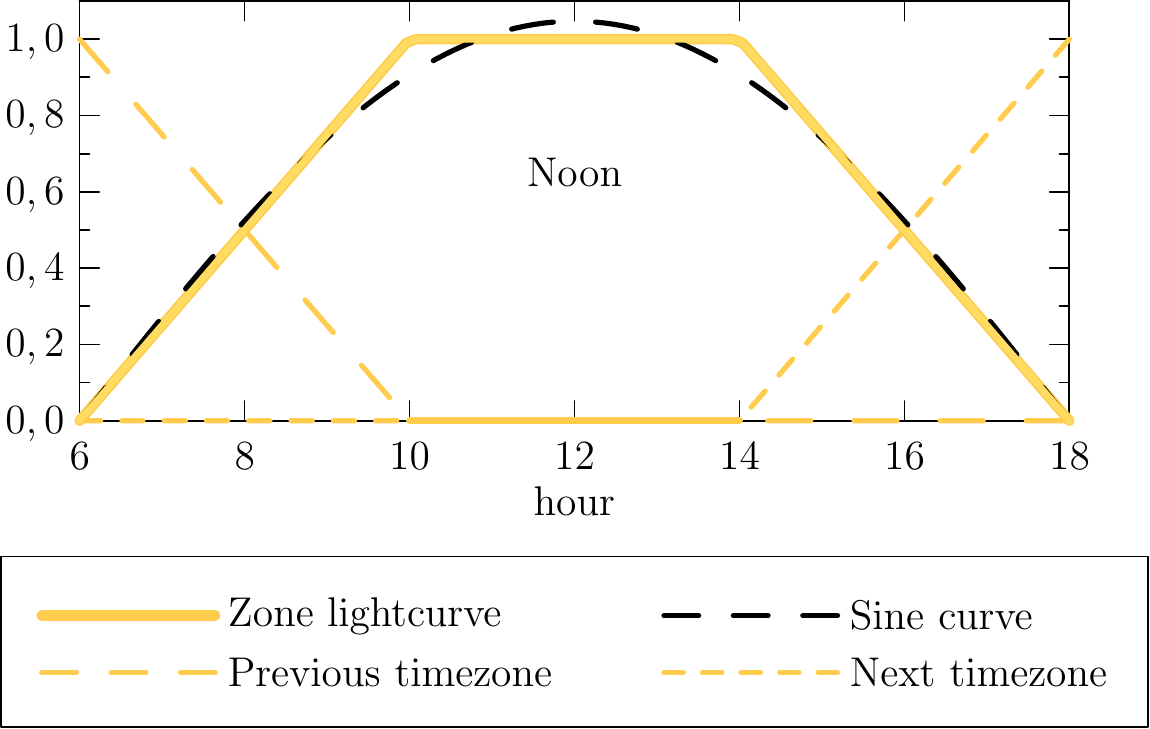}\put(11,58){b)}\end{overpic}
\caption{(a) Behaviour of light in the light channel, (b) the synthetic
  diurnal cycle.}
\label{fig:roof}
\end{figure}

To keep the usage of total incoming sunlight temporally
constant, the rural space is divided into three timezones in the
vertical direction (Fig.~\ref{fig:roof}a). The timezones are shifted
from each other by $\pm 8$ hours. The total amount of light received
by the living space is temporally constant (Fig.~\ref{fig:roof}b). The
light channel has an adjustable ceiling made of (more or less
diffusely) reflecting material. The slope of the ceiling adjacent to a
zone determines the level of sunlight in the zone.

To ensure darkness during simulated night despite some stray light,
blinders are used during nighttime.  In each zone, the sun rises at 6
am and the light level increases linearly until 10 am
(Fig.~\ref{fig:roof}b). The light level stays constant until 2 pm when
it starts to decline linearly towards the sunset at 6 pm. The
piecewise linear light curve approximates the more ideal sinusoidal
dependence rather well (Fig.~\ref{fig:roof}b).

The rural space is 50 m high and 5 km long along the cylinder
axis. There is another upside-down cylinder attached to it which makes
the total length of the unit 10 km. The attached cylinder receives
light from a symmetric light gathering system below.

The light channel is 137 m wide. The width of the light channel
is determined by Sun's angular diameter at Ceres distance.

Adding seasonal variations to the lighting scheme would likely be
possible but it is outside the scope of this paper.

\subsection{Thermal design}

\begin{figure}[ht]
\centering
\includegraphics[width=0.65\textwidth]{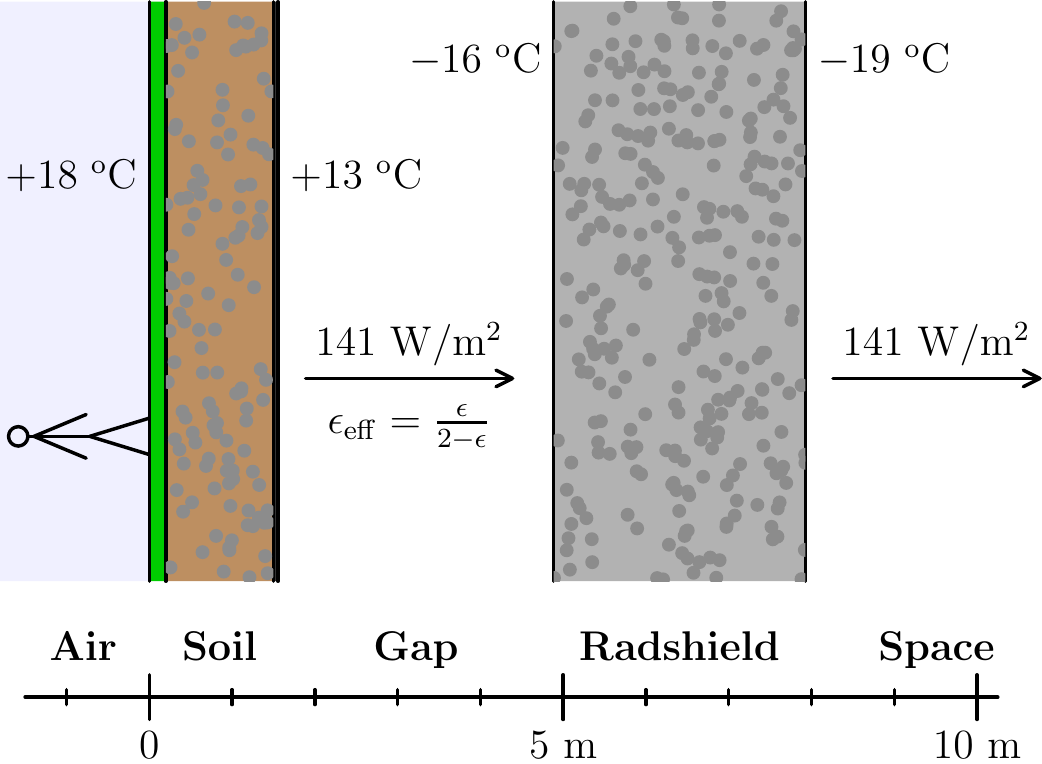}
\caption{Thermal design of the habitat wall. The soil is rotating, the
  radshield is not.}
\label{fig:thermal}
\end{figure}

Minimisation of structural mass implies minimisation of rotating
mass. Therefore we make the radiation shield non-rotating. The shield
must be separated from the habitat by a gap. Figure \ref{fig:thermal}
shows cross-section of the habitat soil and the radiation shield, with
temperatures marked. The heat flux 141 W/m$^2$ matches well with the
wanted insolation of 130 W/m$^2$ plus some extra heat dissipation due
to imperfections in the internal mirrors and dissipation of electric
power consumed inside the habitat.

Both the soil and the radiation shield need an internal liquid heat
transfer system, because the thermal conductivity alone would be
insufficient. In soil, the heat transfer fluid can be water because
all parts are above the freezing point. In the radiation shield the
liquid must be more cold-tolerant, for example some light hydrocarbon
such as heptane. Ceres is rich in carbon and hydrogen for producing
hydrocarbons.

We assume +18 $^\mathrm{o}$C diurnally averaged habitat
temperature. The minimum nighttime temperature is +13 $^\mathrm{o}$C
and the maximum afternoon temperature is +23 $^\mathrm{o}$C. The
dewpoint tends to settle close to the nighttime minimum temperature of
+13 $^\mathrm{o}$C.

For the internal surface adjacent to the vacuum gap, we assume
emissivity of 0.95. For the external surface we assume a slightly
lower emissivity value of 0.92, because it is exposed to potential
modifications caused by space radiation and micrometeoroids.

In the calculation of radiative cooling, we took into account the
effect of the inclined mirrors, using a conservative approximation. We
also took into account the mutual view factors of the habitat
cylinders using a simple ray-tracing calculation for thermal infrared
photons.

We made the calculations for Ceres aphelion. For other phases of the
orbit, some of the mirror elements are turned to reject light.  In
this way, the illumination level inside the habitats and the required
cooling are constant.

\begin{figure}[tbp]
\centering
\begin{overpic}[width=0.85\textwidth]{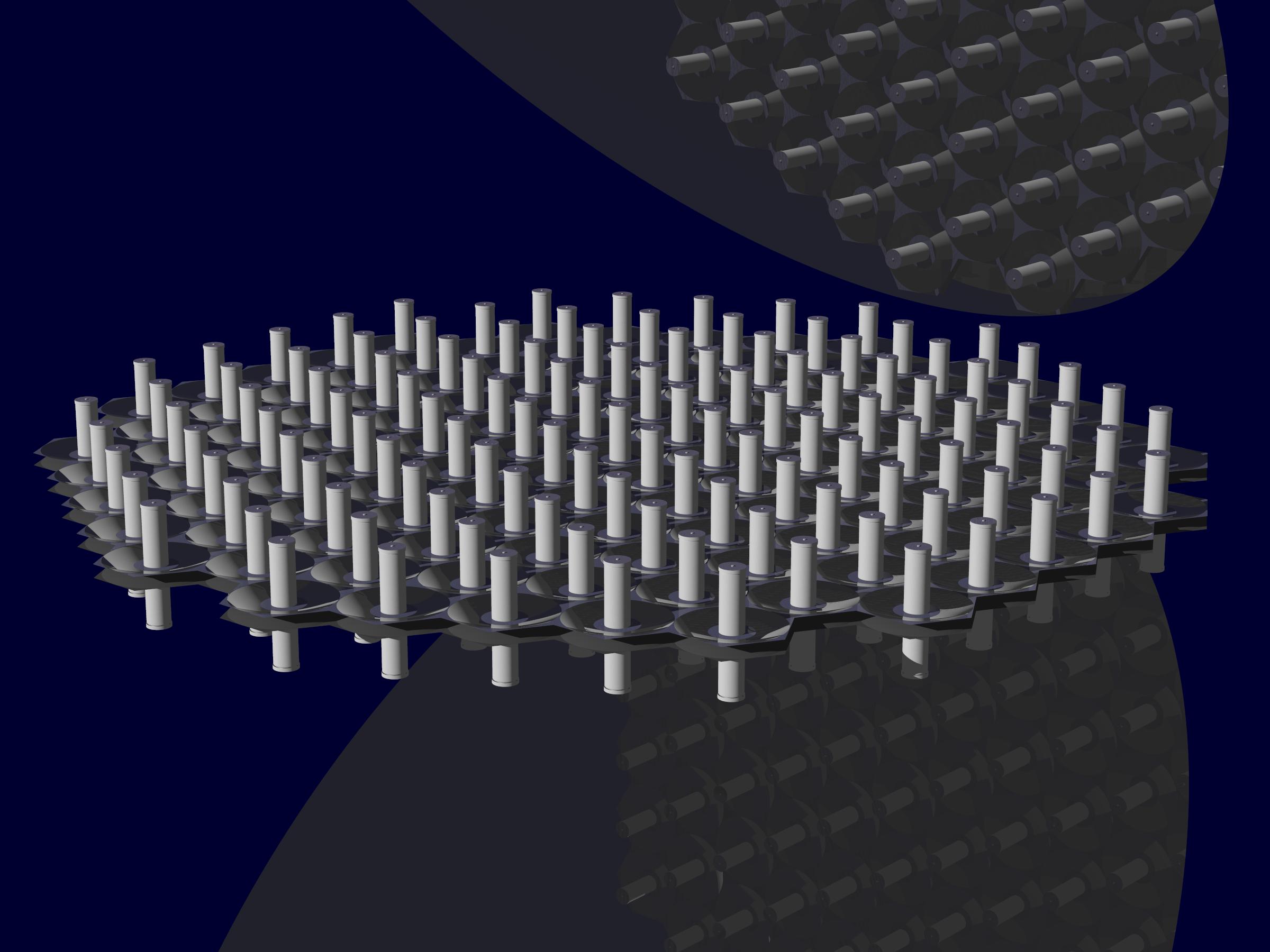}\put(1,70){\textcolor{white}{a)}}\end{overpic}
\vskip 0.5\baselineskip
\begin{overpic}[width=0.85\textwidth]{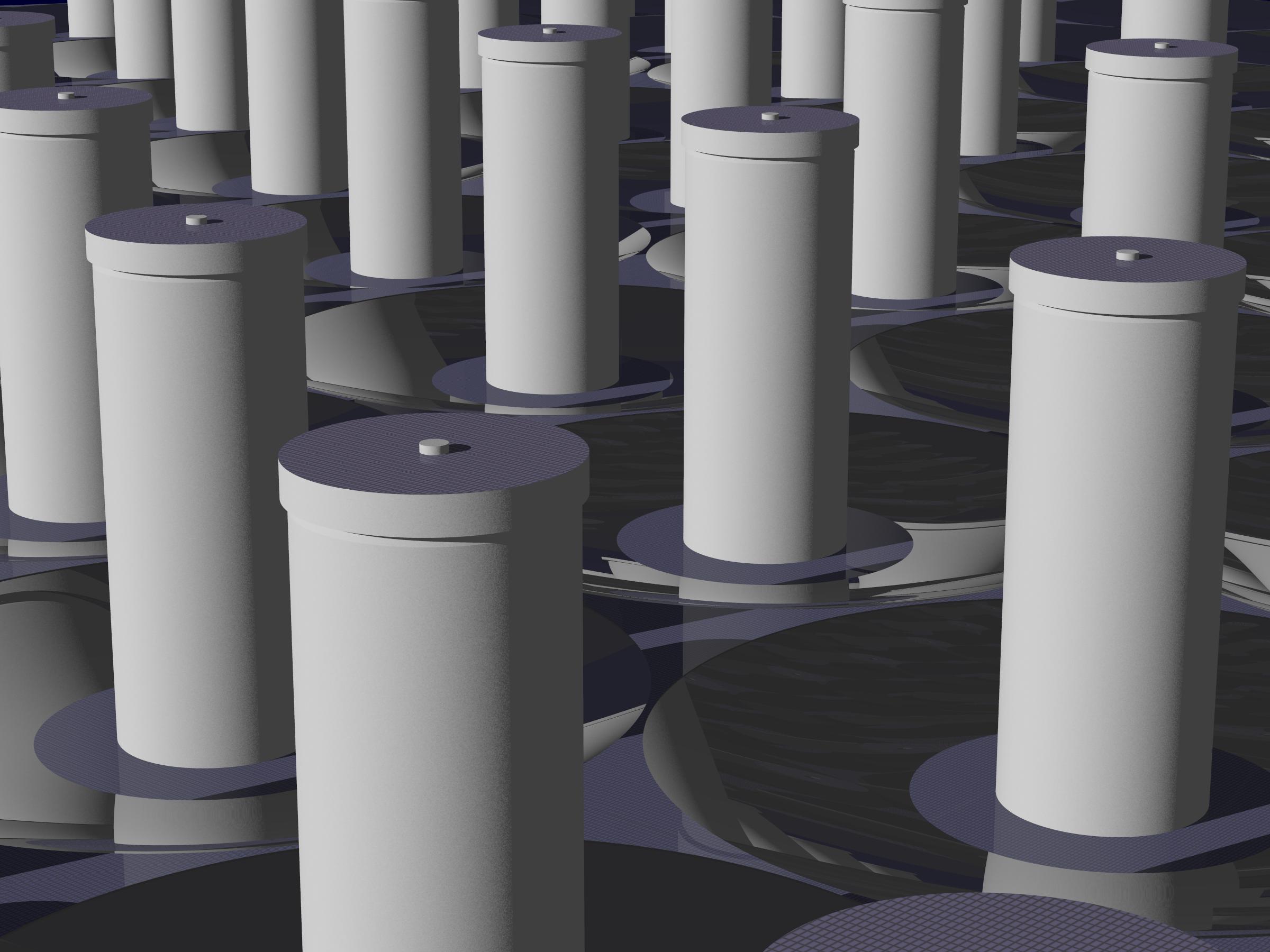}\put(1,70){\textcolor{white}{b)}}\end{overpic}
\caption{(a) The megasatellite, (b) a detail.}
\label{fig:megasat}
\end{figure}

Figure \ref{fig:megasat} shows 3-D renderings of the
megasatellite. The illumination was chosen to help visualisation. We
did not use Sun illumination from a realistic direction because it
would have made the renderings more challenging to understand.

\subsection{Megasatellite frame structural mass}

\begin{figure}[htb]
\centering
\includegraphics[width=0.4\textwidth]{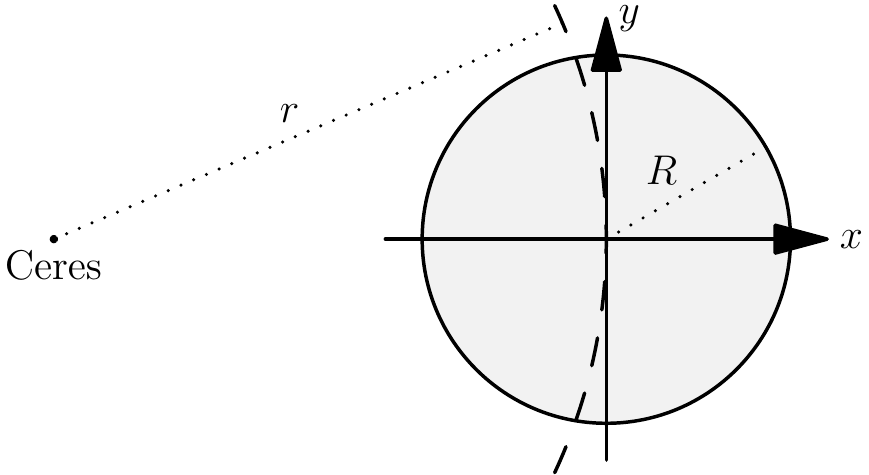}
\caption{Geometry for structural mass estimation.}
\label{fig:aux}
\end{figure}

Consider the megasatellite disk of radius $R$ in equatorial Ceres
orbit with orbital distance $r=10^5$ km (Fig.~\ref{fig:aux}).

When the coordinate origin is fixed at the
centre of the disk, Ceres is at $(x=-r,y=0)$. The gravity field within
the disk along the $x$ axis is
\begin{equation}
g_x = -\frac{G M_C}{(r+x)^2}\,.
\end{equation}
Let us assume $R\ll r$, which is qualitatively valid even for very
large megasatellites e.g.~$R\sim (1/4)r$ and an accurate approximation
for smaller sizes. Then
\begin{equation}
g_x = -\frac{G M_C}{r^2}\left(1+\frac{x}{r}\right)^{-2} \approx 
-\frac{G M_C}{r^2}\left(1-2\frac{x}{r}\right)\,.
\end{equation}
The constant term is cancelled by the orbital centrifugal force so
that the effective gravity within the disk along $x$ ($-R \le x \le
R$) is
\begin{equation}
g_x^\mathrm{eff} = \frac{2G M_C}{r^3} x\,.
\end{equation}
In the same approximation, along the $y$ axis the gravity field is
\begin{equation}
g_y = -\frac{G M_C}{r^3} y\,.
\end{equation}

Consider a uniform column of mass of cross-sectional area $A$ and mass
density $\rho$ within the disk along $y$ from $y=0$ to
$y=R$. The tidal compression force at $y=0$ is
\begin{equation}
F = \int dm g_y = \rho A \int_0^R dy g_y
= \frac{G M_C}{2r^3} \rho A R^2\,.
\end{equation}
The compression is
\begin{equation}
Y = \frac{F}{A} = \frac{G M_C \rho R^2}{2r^3}\,.
\end{equation}
Consider case $R=(1/4)r$ which is a very large megasatellite of
$R=25,000$ km for $8\cdot 10^{11}$ people. If the material is steel
with $\rho=7.8\cdot 10^3$ kg/m$^3$, we obtain
$Y=0.15$ MPa, which on Earth would correspond to steel column height of
\begin{equation}
h_\mathrm{eff} = \frac{Y}{\rho g}
= \frac{1}{2} \left(\frac{G M_C}{r^2 g}\right) \left(\frac{R}{r}\right)^2 r
= 2\,\,\mathrm{m}\,.
\end{equation}
This is quite low structural requirement. If the structural mass
fraction is only 0.1\,\%, for example, then the compression is 150 MPa
which is well tolerated by steel. The tidal pull in $x$ direction is
two times higher.

Overall, the tidal forces increase with the megasatellite radius but
the additional structural mass due to them is only $\sim 0.1$\,\%
even for very large megasatellite of radius $R=r/4=25,000$ km.

\subsection{Mass of magnetic bearings}

The mass of the Inductrack magnetic bearings is many orders of
magnitude smaller than the total mass so it is negligible in the mass
budget. A rough reasoning is that in earthly maglev trains, the
magnetic levitation system is able to carry the weight of the train in
$1 g$ without overwhelmingly dominating the mass budget. In 100,000 km
Ceres orbit, the gravity field of Ceres is $6.4\cdot 10^{-7} g$ and
the microgravity in the satellite is smaller than this by factor $R/r$
where $R$ is the megasatellite radius and $r=10^5$ km is the orbital
radius. Hence the forces that the bearings have to carry are smaller
than on Earth by a factor $\sim 10^7$ or more. Thus the mass fraction
of the magnetic bearings is expected to be $\sim 10^{-8}$ or less.

\subsection{Interconnectivity}

Fast and easy travel between the habitats is necessary. In order to be
long-term sustainable, the travel must also be propellantless. It
cannot be based on rocketry.

A straightforward way is that the passenger stations are located at
the ends of the rotating habitat cylinders, at the cylinder
axes. Passengers use an elevator to get to the axis. They experience
weightlessness during the trip, apart from the acceleration and
deceleration phases of the vehicle when it moves in a tunnel that
connects the habitats. There are many technical options for the
vehicles. The vehicles might resemble autonomous cars or they might be
more like trains. They might move on wheels along rails or use
magnetic levitation. They could move in vacuum or in air. If they move
in vacuum, either the vehicles or the passengers must move through
airlocks at both ends of the trip.

When first encountered, weightlessness causes nausea and vomiting for
some people. However, in a settlement where people experience
occasional weightlessness from childhood, it is plausible to think
that they can tolerate it well during short trips. Nevertheless, since
this not known with certainty, travel solutions where weightlessness
can be avoided are worth studying. Gravity can be maintained at the
stations if the station is placed at the perimeter of the
cylinder. Vehicles move at constant speed, which is the same as the
rotation speed at the distance from the centre where the station
is. The vehicles do not stop at the station because the station is
moving at the same speed as the vehicle. One can have gravity also
during the trip if one winds the tunnels into spirals. This mode of
transportation can coexist with the baseline zero gravity version
explained above, because the stations are located in different areas
of the habitat. Having more than one mode of transportation available
increases safety by providing redundancy. As a safety feature for
abnormal situations, there must also be a docking port for external
spacecraft in both ends of each habitat cylinder.

The tunnels probably do not need radiation shielding against GCRs,
because people spend only small fraction of their time
travelling. Full shielding of the tunnels would increase megasatellite
mass by only 2--3\,\%, though.

\subsection{Geometric parameters and mass budget}

Table \ref{tab:geom} lists the main parameters. The
parameters are listed for the twin cylinder that goes through the
symmetry plane and comprises also the downward pointing parts seen in
Fig.~\ref{fig:megasat}a. The two halves of the habitat cylinder are
physically connected and enclosed into the same radiation shield.

\begin{table}[htb]
\caption{Geometric parameters of the twin cylinder habitat.}
\begin{tabular}{lll}
\hline
Habitat radius $R_H$ & 1 km \\
Habitat length       & 10 km \\
Parabolic reflector radius     & 4.4 km \\
Population           & 56700 \\
Living area          & 114 km$^2$, 2000 m$^2$/person, 44.9 \% urban \\
Urban gravity        & 81 \% \\
Electric power       & 6.26 kW/person, 7.0 W per urban m$^2$ \\
Rotation period      & 1.06 min \\
Power dissipation    & 156 kW/person \\
Heat flux            & 141 W/m$^2$ \\
Light channel width  & 137 m \\
Light channel insolation & 5.0 suns \\
\hline
\end{tabular}
\label{tab:geom}
\end{table}

The electric power given in Table \ref{tab:geom} was reached in the
following way. We assumed that all areas not covered by the paraboloid
collector (the habitat cylinder roof, a ring around the cylinder, and
areas between adjacent paraboloids) are used for solar panels that
have 20\,\% efficiency. Further, we made arbitrary assumption that
50\,\% of the produced power is used for the habitats while the other
half is reserved for megasatellite level functions such as the
transportation network and attitude and orbit control.

Table \ref{tab:massbudget} gives the mass budget of the megasatellite
as computed from the geometric model using parameters of Table \ref{tab:geom}.

\begin{table}[htb]
\caption{Mass budget of megasatellite.}
\begin{tabular}{lrr}
\hline
Item & Mass/person & Fraction \\
\hline
Shield & 6712 t & 68.7 \%\\
Soil and biosphere & 2481 t & 25.4 \%\\
Structural & 484 t & 5.0 \%\\
Air & 96 t & 1.0 \%\\
\hline
Total & 9774 t & 100 \%\\

\hline
\end{tabular}
\label{tab:massbudget}
\end{table}

\subsection{Manufacturing energy}

The manufacturing energy is dominated by making tensile structural
material, which is needed to contain the centrifugal force of
artificial gravity and the internal pressure. One option for the
structural material is piano wire steel (music wire, ASTM A228 alloy,
0.7--1\,\% C, 0.2--0.6\,\% Mn, remainder Fe) whose yield strength is
listed by MatWeb as 1590--1760 MPa. In the calculation we have
assumed tension of 760 MPa which is a bit less than half of the lowest
table value.

In 1998, \citet{DeBeerEtAl1998} estimated the worldwide average energy
consumption for steel making to be 24 MJ/kg, while the most
energy-efficient process used 19 MJ/kg and reductions to 12.5 MJ/kg
were foreseen, or down to 10 MJ/kg with waste heat recovery
techniques. In this paper we adopt the value of 20 MJ/kg for steel and
ignore the energy consumption of other than structural materials.

\subsection{Impact threat}
\label{subsect:impactthreat}

Impacts are a potential new threat. However, this risk can be
effectively mitigated or perhaps even eliminated
completely. Collisions of sub-meter scale meteoroids do not penetrate
the radiation shield. For large impactors, early detection and evasive
manoeuvre of the habitat or deflection of the impactor is feasible. If
there remains a class of mid-range impactors that is not covered by
these strategies, one can use active near-range defence based on
kinetic kill vehicles or/and lasers.  Finally, if warning comes too
late to apply other defences, one can evacuate people from the
threatened end of the cylinder habitat, or if there is sufficient
time, from that cylinder altogether.

\section{Obtaining the materials}

\subsection{Space elevator}

\begin{figure}[ht]
\centering
\includegraphics[width=0.3\textwidth]{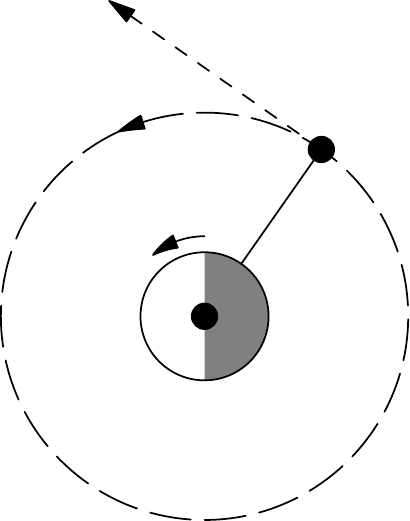}
\caption{Ceres space elevator.}
\label{fig:elevator}
\end{figure}

The Cererian equatorial escape speed is only 426 m/s when launching
along rotation. Because Ceres has water, one could electrolyse water
and lift material by H$_2$/O$_2$ rockets. However, energetically it is
much cheaper to use a space elevator (Fig.~\ref{fig:elevator}). The
altitude of the stationary orbit is 722 km and the altitude of the
tether tip to inject to elliptical orbit reaching 100,000 km altitude
is 1024 km. Thus the tether length is 1024 km. For example, if the
tether is made of dyneema (ultra-high molecular weight polyethylene
fibre, mass density 975 kg/m$^3$) tensioned at 1240 MPa (40\,\% of the
tensile strength of 3.1 GPa), the tether mass is only 23\,\% of the
lifted payload mass. If the lifting speed is 10 m/s, for example, the
lifting takes 28 hours. A round cable tether might break due to a
meteoroid impact but a non-planar tape tether or tubular tether
resists impacts well.

A 5 cm wide and 25 $\mu$m thin initial dyneema tether weighs 1.25
tonnes. It can lift 5000 kg at one time. If the lifting speed is, for
example, 10 m/s, lifting takes 28 hours and the average power
consumption is 2.6 kW. Thus, an initial tether which is sufficiently
wide that it does not break due to micrometeoroids can be installed by a
spacecraft of a few tonnes mass. The initial tether will
be later replaced by a thicker one which was manufactured from the
lifted materials, and the process is repeated. The lifting capacity
grows exponentially because in each step, the new tether is many times
thicker than the old one.

After releasing from the tip, the payload enters an elliptic orbit
which must be circularised by a 20 m/s apoapsis propulsive burn. For
example, with specific impulse of 100 s (steam rocket), the propellant
fraction is 2\,\%. Per mass unit, the energy needed by the elevator is
54 kJ/kg. Making +300 C steam from -130 C Cererian ice requires $\sim
4$ MJ/kg of heat energy, which multiplied by the propellant fraction
yields 80 kJ/kg. Thus, in total the energy needed to lift from surface
to 100,000 circular orbit is 54+80=134 kJ/kg. If using a H$_2$/O$_2$
rocket, the energy cost would be $\sim 20$ times higher.

\begin{figure}[ht]
\centering
\includegraphics[width=0.6\textwidth]{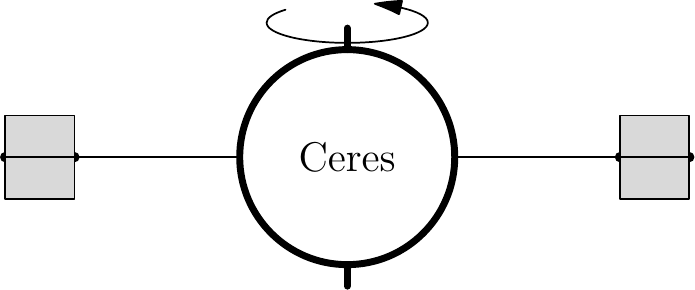}
\caption{Large twin space elevator system with solar panels.}
\label{fig:2elev}
\end{figure}

The number of elevators could be more than one. For example, Figure
\ref{fig:2elev} shows two elevators on opposite sides of Ceres.  Their
combined area of solar panels produces average power which is
equivalent to covering 20\,\% of Ceres surface by solar panels.

\subsection{Surface transportation on Ceres}

On the surface of Ceres, the mined material must be transported to the
elevator(s). The transportation can use wheeled trucks. The trucks can
have large solar panels to generate power. Large panels are feasible
because the gravity is only $(1/34) g$. On Ceres, the night lasts only
4.5 hours. Therefore the vehicles, once they are of sufficient size,
can stall during the night and rely on insulation and thermal inertia
to maintain sufficient internal temperature. The vehicles do not
necessarily need energy storage devices. This is a welcome
simplification because the energy storage devices would have to be
ISRU.

\section{Resource limits}

The above analysis can be summarised by the following numbers:
\begin{itemize}
\item Mass per living area 5000 kg/m$^2$, of which 5\,\% (250 kg/m$^2$) is
  structural mass (steel) whose production takes 20 MJ/kg of
  energy.
\item Thus, production energy is 1 MJ/kg. Per living area it is $5 \cdot 10^9 $J/m$^2$.
\item Living area 2000 m$^2$/person, thus $10^7$ kg/person and
  $10^{13}$ J/person.
\item Megasatellite disk area 1175 m$^2$/person.
\end{itemize}

\subsection{Ceres mass resource}

Ceres mass $M_C=9.38\cdot 10^{20}$ kg $\approx 10^{21}$
kg would suffice for building 200 billion square kilometres
of living area, i.e., 400 times larger than Earth's surface area. At
population density of 500 /km$^2$, such area would accommodate
$10^{14}$ people, i.e., $10^4$ times current world population. A
megasatellite of this extreme size would reside in a heliocentric
orbit, because it would not fit in Ceres orbit.

\subsection{Ceres angular momentum resource}

Ceres rotation period is 9.07 hours and radius $R_C=470$ km. The internal
mass distribution is not known but let us assume for simplicity that it is
uniform. Then the moment of inertia is
\begin{equation}
I_C=\frac{2}{5}M_C R_C^2=8.3\cdot 10^{31}\, \mathrm{kg}\, \mathrm{m}^2
\end{equation}
and the angular momentum is $L_C=1.6\cdot 10^{28}$ Nms. At tether tip
at 1024 km altitude, the tip speed is 287 m/s so the angular momentum
per injected payload mass is $4.3\cdot 10^8$ Nms/kg. Then each kilogram
of launched material reduces Ceres angular momentum by fraction $2.7\cdot 10^{-20}$.
For example, launching $10^{18}$ kg of material reduces Ceres angular
momentum by 2.7\,\%. This amount suffices to
build $200 \cdot 10^6$ km$^2$ of living area for $10^{11}$
people. The megasatellite's diameter is 12,000
km, i.e.,~12\,\% of the orbiting distance.

If there is a need to go beyond the angular momentum resource of
Ceres, one could extend the elevator tether and compensate for the
angular momentum loss by applying propulsion at the extended tip. The
energy needed by propulsion is independent of the tether length, while
the needed propellant mass decreases if the tether is made longer. The
energy needed is a few times larger than with plain space elevator
but significantly less than lifting material by H$_2$/O$_2$ rocket
propulsion. The details are given in Appendix A below.

The internal structure of Ceres is poorly constrained at the moment.
If the structure is such that the crust and the core are separated by
a fluid layer, then speculatively it might happen that using the space elevator
would cause the crust to slow down relative to the core.
If the resulting velocity shear in the fluid layer is
concentrated to small regions, those regions might heat up sufficiently
to potentially cause cryovolcanism, which would be an undesired side effect.  In such case, one might apply the
angular momentum restoration strategy mentioned in the above
paragraph already earlier. There is no risk that such situation would
develop unnoticed, because monitoring the rotational response of the
surface to space elevator operation reveals if velocity shear is developing below.

\subsection{Speed of lifting material}

If one covers 20\,\% of Ceres area by solar panels, or places an
equivalent power production capacity in the space elevator(s) directly
(Fig.~\ref{fig:2elev}), one produces 4.8 TW of electric power. Lifting
material by the elevator takes 54 kJ/kg. One also needs 20 m/s of
propulsive delta-v to circularise the orbit, and to produce the
propellant also takes energy. Let us assume, however, that the orbit
circularisation energy is generated during the 52-days long trip of
the payload from the elevator tip to the apoapsis.

We then obtain that one can lift $2.8\cdot 10^{15}$ kg/year. This
suffices for 280 million new people per year. This is larger than
current yearly increase of world population, which is about 80 million
per year. A large megasatellite of $10^{11}$ population and $200\cdot
10^6$ km$^2$ living area takes 360 years to build at this rate.

\subsection{Timescale of energy-limited bootstrapping}

Let us consider the question in what timescale the infrastructure
could be bootstrapped. The space elevator can lift its own mass in
only few hours. From the lifted material, a larger power system and
thicker tether must be made. Because lifting mass is fast, mass does
not limit the growth rate. The limitation comes from how fast the
existing power system can generate the amount of energy that is needed
to produce the next generation power system from the lifted
material.  Going from an initial kilowatt level to final terawatt
level takes $\sim 30$ doublings. If the doubling time (the energy payback
time) of the power system is, for example, 4 months, then
bootstrapping takes 10 years. The fact that Ceres is at larger
heliocentric distance does not increase the payback time because it
can be compensated by reflecting concentrators. In the Cererian orbit,
microgravity conditions prevail. In microgravity, parabolic
concentrators are energetically cheap to make because they can be
lightweight.

Because the growth is not mass-limited, the elevator can produce mass
for bootstrapping of the megasatellite, simultaneously with
bootstrapping itself. First one builds the megasatellite with frame,
parabolic concentrators and solar panels, but without the habitats and
radiation shields. This is done to scale up the power production
capacity as quickly as possible. When the power production capacity is
ready, one starts to produce structural material for the habitats, which
takes the majority of the energy. Per living area, $5\cdot 10^9$ J/m$^2$ is
needed, while the insolation on living area is 130/2=65 W/m$^2$ (rural
insolation 130 W/m$^2$ divided by 2 because living area includes also
the urban space which is roughly 50\,\%). Assuming 20\,\% solar panel
efficiency, we thus produce $0.2\times 65$ = 13 W/m$^2$ of
power. Dividing $5\cdot 10^9$ J/m$^2$ by 13 W/m$^2$ yields 12 years
timescale for making the rest of the habitat, once the power system is ready.

\subsection{Overall bootstrapping timescale}

In summary, if a power system's doubling time (payback time) is 4 months,
bootstrapping the elevator and the power system of the megasatellite
takes 10 years. Then, using the megasatellite's power system, it takes
another 12 years to build the habitats and the radiation shields, with
total time of 22 years. In addition, the area of solar panels that
fits in Ceres stationary orbit limits the building speed to 280
million new people per year, if panel area equivalent to 20\,\% of
Ceres surface is used as illustrated in Fig.~\ref{fig:2elev}.

Our bootstrapping analysis concerned only the physical limits of mass,
energy and angular momentum. Physics allows
bootstrapping in only 22 years under the stated assumptions. It is likely,
of course, that the actual bootstrapping timescale would be longer, or
much longer, and instead of physics, it would be driven by
technological delays and the logistics of transporting the inhabitants
from Earth.

\section{Discussion}

Interestingly, cooling is a more important design driver than solar
power. For example, cooling restricts the number of concentric rural
spaces to one. If the heliocentric distance would be smaller, the
habitat cylinders would be closer together since the paraboloid
mirrors would be smaller. This would increase the mutual view factors
between the habitats and thereby reduce cooling. It seems that Ceres
is just at the right heliocentric distance to make the symmetric
two-sided mirror architecture feasible without having to reject
sunlight.

Our cooling system uses also liquid heat transfer, but only over a few
metres across the soil and the radiation shield. Then the liquid heat
transfer system can consist of many small loops so that if a pipe
breaks for any reason, the amount of fluid released is small.

There is a tradeoff between soil thickness and the energy cost of
manufacturing. The default soil thickness of 1.5 m can be increased up
to 4 m, at the cost of refining more of the Ceres raw material into
tensile structural material such as steel or dyneema. The total mass
would not change, only the energy expended in processing the raw
material of Ceres soil would be linearly higher.

The goal is to be able to use Ceres material in sequential order
without sorting out any of it. The goal seems feasible, because the
radiation shields and soil are the largest mass components and their
composition is flexible. The radiation shields and soil are
effectively our trash bin. The iron mass fraction of Ceres is 13--17
\% \citep[Fig.~7]{PrettymanEtAl2019} so there is enough raw material
for structural steel. The only thing that might spoil this plan is if
Ceres turns out to have less than $\sim 0.75$\,\% of nitrogen. In that
case, mining of nitrogen might be necessary from deep subsoil fluids
of Ceres, for example, or one would have to reduce the 50+15 m
atmospheric height.

The megasatellite is growable from one to millions of habitats. The
megasatellite is grown simply by building new habitats at the edges and
expanding the mirrors correspondingly
(Fig.~\ref{fig:clamshellgrowth}). As the megasatellite grows, at some
point one has to add some structural reinforcement also to existing
portions of the megasatellite, but mass-wise the additions are
minor. Also the traffic system may have to be upgraded because growth
of the world increases the amount of through traffic.

The utility value of the megasatellite becomes apparent if we compare
it with traditional surface settlements. It would be technically
possible to colonise the surface of Ceres by centrifuge
habitats. However, then the magnetic bearings would have to carry the
weight of the habitat. The weight is 34 times less than on Earth but
many orders of magnitude more than in the megasatellite's microgravity
conditions. Another challenge would be solar illumination. Because
Ceres rotates, the solar concentrator mirrors would have to track the
Sun, and the 9-hour diurnal cycle does not suit people's natural daily
rhythm. To avoid these issues, probably the habitats would be
artificially illuminated. The small size of Ceres would limit the
available solar power, which would limit the population to few hundred
million. Even if there would be a non-solar energy source, the surface
area of Ceres would limit the total power. The megasatellite's area
can grow much larger than Ceres. Because the Sun shines without
interruptions and always from the same direction, it is feasible to
use sunlight directly for illumination. This is aesthetically pleasing
and energetically efficient.

The megasatellite has two superficial drawbacks relative to surface
settlements. One is that one must build the megasatellite frame and
the mirrors. However, their mass fraction is 1 per cent or less. The
other drawback is that materials must be lifted to orbit, but this is
not significant because the lifting takes only 134 kJ/kg of
energy. This is much less than the 1 MJ/kg needed to process them. The
processing energy is dominated by the tensile structures needed to
create artificial gravity and pressure containment, which are needed
in surface dwellings as well. Furthermore, it is cheaper to obtain the
energy in high orbit where there is no eclipsing and the Sun shines
from a constant direction so that parabolic concentrators become
simple. The concentrators are also lightweight because structurally
they need to withstand only microgravity, not Ceres gravity.

The scale (radius, length) of the habitats is a free parameter. If the
habitats are made larger, the manufacturing energy increases roughly
linearly with radius, because one needs more structural material to
withstand the centrifugal force. The total mass does not change. Also,
nothing prohibits having habitats of different size in the same
megasatellite.

The overall level of difficulty of executing this project is
probably similar to settling Mars. The delta-v and triptime to Ceres
are longer, but on the other hand one avoids planetary landings and
the atmospheric weather and dust. On Ceres it requires some effort
to lift the materials to orbit using the elevator, but it is
energetically cheap. Once the materials are in high Ceres orbit, the
thermal environment is uniform and energy is easy to get due to
absence of eclipses.

\section{Conclusions}

We have analysed a megasatellite settlement that provides $1 g$ gravity,
interconnectivity, growability to beyond current world population,
2000 m$^2$ living area per person, thick soil to enable a natural
environment with trees, along with optimal weather and absence of
natural disasters. The settlement is built in Ceres orbit by lifting
Ceres materials by a space elevator. The mass per person is $10^7$ kg
and the manufacturing energy per person is $10^{13}$ J. The mass is
dominated by radiation shields and soil. The manufacturing energy is
dominated by tensile structural material such as steel needed to
withstand the centrifugal force of artificial gravity and the pressure
containment.

We select Ceres as the source body because it is more likely than
C-type asteroids to have sufficient nitrogen.  Nitrogen is a critical
element because it is needed for the settlement atmospheres. We use
a disk geometry for the megasatellite because its symmetry eliminates
tidal torque so that reaction wheels are not needed to maintain
attitude. The habitats are illuminated by natural sunlight. The
sunlight is gathered onto the disk by two planar mirrors inclined at
$45^\mathrm{o}$ angle and concentrated to desired intensity by
parabolic mirrors.

Future work should involve analysis of the exponential bootstrapping
phase and transportation of people from Earth to Ceres. Also,
different technical alternatives for the travel system between the
habitats should be looked into. Orbit simulations are needed to find
what is the highest altitude of a long-term stable orbit around Ceres.

\section*{Acknowledgements}

The results presented have been achieved under the framework of the
Finnish Centre of Excellence in Research of Sustainable Space (Academy
of Finland grant number 312356). The author thanks Kadri Bussov for
discussions that led to changing the geometry and Sini Merikallio and
Al Globus for commenting the manuscript.

\clearpage
\appendix

\setcounter{figure}{0}

\section{Preventing angular momentum loss by propulsion}

\begin{figure}[htb]
\centering
\includegraphics[width=0.55\textwidth]{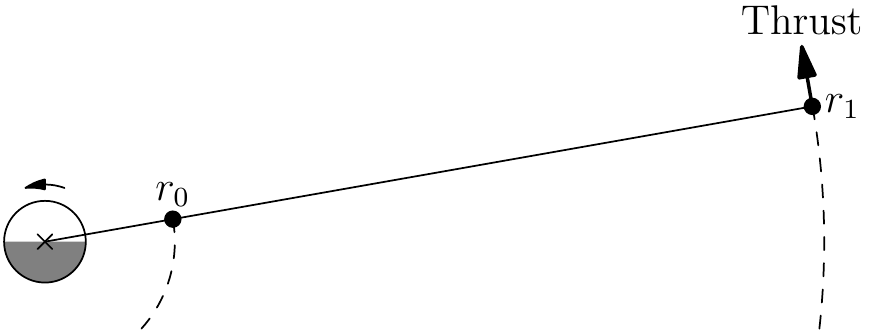}
\caption{Geometry of the calculation of the Appendix.}
\label{fig:appA}
\end{figure}

\subsection*{Nomenclature of the Appendix}
\nobreak\noindent
\begin{longtable}{ll}
$E$          & Energy of burn \\
$k$          & Numeric coefficient \\
$L$          & Generic angular momentum \\
$L_m$        & Angular momentum due to lifting of mass $m$ \\
$m$          & Mass of lifted propellant \\
$M$          & Mass of lifted payload \\
$r$          & Orbit radius, $10^5$ km \\
$r_0$        & Radial distance of tether tip \\
$r_1$        & Radial distance of tip of tether extension \\
$R_C$        & Radius of Ceres, 470 km \\
$v_0$        & Thruster exhaust speed \\
$\eta_T$     & Efficiency of the thruster \\
$\omega_C$   & Ceres angular rotation speed, $1.923\cdot 10^{-4}$ s$^{-1}$ \\
\end{longtable}

Consider some payload mass $M$ lifted from Ceres surface to the
elevator tip at radial distance $r_0 = R_C+1024$ km from Ceres'
centre, Fig.~\ref{fig:appA}. The tether length 1024 km is sufficient
to inject objects to elliptic orbit whose apoapsis is at $r=10^5$ km
distance. The angular momentum loss of Ceres is
\begin{equation}
L = M (r_0^2-R_C^2) \omega_C\,.
\end{equation}
Assume that there is an extension of the tether up to some larger
distance $r_1 > r_0$. We apply propulsion at $r_1$ to try and cancel the
angular momentum loss. The propellant mass $m$ is transferred from
Ceres along the tether. After lifting it to $r_1$, its
angular momentum has increased by $L_m=m (r_1^2-R_C^2) \omega_C$. This angular
momentum was also taken from Ceres, so overall the propulsive burn must create angular
momentum $L+L_m$. The exhaust speed $v_0$ of the thruster
(i.e., specific impulse in velocity units) must be at least $r_1 \omega_C$.
Let us write $v_0=k r_1 \omega_C$ where $k>1$. The produced angular
momentum is $m r_1 v_0$. Equating it with $L+L_m$ and solving for $m$ yields
\begin{equation}
m = M \frac{r_0^2-R_C^2}{(k-1) r_1^2 + R_C^2}\,.
\end{equation}
The energy consumed by the burn is
\begin{equation}
E = \frac{1}{2\eta_T} m v_0^2
\end{equation}
where $\eta_T$ is the efficiency of the thruster. Let us select, for
example, $k=2$. Typically $r_1^2 \gg R_C^2$, and using this
approximation we obtain
\begin{equation}
E \approx \frac{2}{\eta_T} M (r_0^2-R_C^2)\omega_C^2
\end{equation}
which is independent of $r_1$. For example, if $r_1=6 r_0 = 8964$ km
and thrust efficiency $\eta_T=0.5$, then $v_0=3.45$ km/s, $m/M = 0.025$
and $E/M = 300$ kJ/kg, which is 5.6 times larger than the energy
needed to lift mass $M$ to the normal elevator tip $r_0$. In principle one
could get back part of this energy because lifting $m$ to $r_1$
liberates rather than consumes energy.

In summary, it is possible to prevent Ceres angular momentum loss by
using propulsion at tip of elevator tether extension.  The energy
expended in making the propellant is nearly independent of the tether
length.  The propellant mass is reduced if one makes the tether
longer. The specific impulse is proportional to the tether length,
although there is some freedom. Chemical hydrogen-oxygen thruster is
among the possibilities.  It implies $\sim$ 10,000 km tether length
and $\sim 2.5$\,\% of the lifted mass being water which is
electrolysed and consumed as propellant. The energy and propellant
mass consumed by this scheme is less than lifting the material from the
surface by direct chemical propusion.






\end{document}